\documentstyle[twocolumn,aps,prl,epsf,psfig,epsfig,pstricks,amssymb]{revtex}
\input{epsf}
\input epsf.sty
\begin{document} 

\wideabs{
\title{Similarities between  the Hubbard and Periodic
Anderson Models\\ at  Finite Temperatures}
\author{K. Held,$^{1}$,
C. Huscroft,$^{2}$ R.T. Scalettar,$^3$ and A.K. McMahan$^4$}
\address{$^1$Theoretische Physik III, Universit\"at Augsburg,
86135 Augsburg, Germany}
\address{$^2$Physics Department, University of Cincinnati, 
Cincinnati, OH 45221-0011}
\address{$^3$Physics Department, University of California, Davis, CA 95616}
\address{$^4$Lawrence Livermore National Laboratory, University of California,
Livermore, CA 94550}
\date{\today}

\maketitle
\begin{abstract}
The single band Hubbard and the two band 
Periodic Anderson Hamiltonians have traditionally been 
applied to rather different
physical problems -  the Mott transition and itinerant magnetism,
and Kondo singlet formation and scattering off localized magnetic states,
respectively.  In this paper, we compare the magnetic and charge
correlations, and spectral functions, of the two systems.
We show quantitatively that they exhibit remarkably similar behavior, 
including a nearly identical topology of the 
finite temperature phase diagrams at half-filling.
We address potential implications of this for theories
of the rare earth ``volume collapse'' transition.\\

\end{abstract}


}

The single band Hubbard model (HM) and the two band Periodic Anderson model
(PAM) have a long history in the study of 
many body effects in solids.
While they have been proposed to describe
complex d-  and  f-electron systems, respectively, 
the detailed atomic physics and orbital structure are
nevertheless eliminated, leaving minimal Hamiltonians 
where the interplay between
electron kinetic energy,
screened electron-electron repulsion, the Pauli principle, 
temperature, and electron density
still give rise to a rich variety of phenomena.

The key features of the half-filled HM are antiferromagnetism and a
Mott insulator-metal transition within the paramagnetic phase
\cite{GEBHARD,DMFT}.  Originally
\cite{HUBBARDIII}, the Mott transition was associated with a change in
the density of states (DOS) from widely separated upper and lower Hubbard
bands to a broad one-peak structure, and an accompanying loss of local
moments. The key features of the PAM are magnetism
arising from an indirect `RKKY' interaction of the local moments
mediated by the conduction electrons,  and a DOS which,
in the paramagnetic phase, has a three peak structure - upper and
lower bands and an additional Abrikosov-Suhl resonance near the Fermi
surface\cite{NEWNS}. Concomitant with the development of this resonance,
local moments become screened by the conduction electrons, and
an anomalous sharp feature in the free energy is observed \cite{HUSCROFT}. 
During the past years, it has been
established by dynamical mean-field theory (DMFT) \cite{DMFT,DMFT2} and 
quantum Monte Carlo (QMC) simulations \cite{QMCHM}
that the HM also has a three peak
structure in its paramagnetic DOS, thereby blurring at
least one of the features believed to distinguish the two models.

It is the purpose of this paper to note even greater similarity between
the two models at finite temperature, especially if a more physical
choice is taken for the interband hybridization in
the PAM.  
Specifically, our new DMFT results show that the f-electrons of the
PAM undergo a transition similar to the Mott transition 
of the HM, as is reflected by the f-electron spectral function, 
quasiparticle weight, and charge compressibility.
Furthermore, the phase diagrams,
at finite temperature,
associated with the two models are topologically equivalent.

   These similarities are of particular interest in that the HM
   \cite{JOHANSSON} and PAM (or its impurity approximation
   \cite{ALLEN}) have been proposed as competing models for the
   ``volume-collapse'' transitions in the rare earth and actinide
   metals.  These pressure-induced phase transitions are characterized
   by unusually large (9--15\%) volume changes, and are believed to
   arise from f-electron correlation effects \cite{BENEDICT,JCAMD}.
   The present paper suggests that the predictions of the two models
   are more similar than has previously been thought, and that
   perceived incompatibilities may have arisen in part from differing
   and less rigorous approximations used in the treatment of electron
   correlations, as for example in mean-field-like local-density
   functional theory \cite{JCAMD}.

The single band HM is given by,

\begin{equation}
H = 
\sum_{k \sigma} \epsilon^f_k \;
f_{k\sigma}^{\dagger} f_{k\sigma}^{\phantom{\dagger}} \nonumber\\
+ U_{f} \sum_{i} (n_{if\uparrow} - \frac12)
(n_{if\downarrow} - \frac12).
\label{hmham}
\end{equation}   
Here $\epsilon^{f}_{k} \!=\! 
\epsilon^{f}_0 -2 t_{ff} [ \cos(k_x)\!+\!\cos(k_y)\!+\!\cos(k_z)  ]$ is the 
nearest-neighbor band dispersion,
where the energy scale is set by $t_{ff}=1$,
and $U_{f}$ is the f-electron local interaction on every lattice site $i$.
The two band PAM is,

\begin{eqnarray}
  H &=& \sum_{k \sigma} [ \epsilon^d_k\;
  d_{k\sigma}^{\dagger} d_{k\sigma}^{\phantom{\dagger}}
  +  \epsilon^f_k \;
  f_{k\sigma}^{\dagger} f_{k\sigma}^{\phantom{\dagger}} 
  + V_k\;
  (d_{k\sigma}^{\dagger} f_{k\sigma}^{\phantom{\dagger}}
  + h.c.)]\nonumber \\ && + U_{f} \sum_{i} (n_{if\uparrow} - \frac12)
  (n_{if\downarrow} - \frac12),
\label{pamham}
\end{eqnarray}   
where  $V_k$ denotes the hybridization between f and d electrons
at momentum $k$.
In the PAM, the d-electrons are chosen to have the same nearest-neighbor
dispersion as the f-electrons of the HM ($t_{dd}=1$), 
while the f-electrons are chosen to be dispersionless, $t_{ff}=0.$
We study the ``symmetric'' PAM, i.e.,
$\epsilon^{d(f)}_0=0$ and chemical potential $\mu=0$, 
which has the property that
$\langle n_f \rangle= \langle n_d \rangle= 1$ for
all choices of $U_f$, $t_{dd(ff)}$, $t_{fd}$, and temperature $T$.
Finally, we choose a nearest-neighbor intersite hybridization
$V_{k}\! =\! -2 t_{fd}\; [ \cos(k_x)\!+\!\cos(k_y)\!+\!\cos(k_z)  ]$, 
but also have investigated the more canonical onsite hybridization
$V_{k} \!=\! \sqrt{3} t_{fd}$, as well as an odd-parity nearest-neighbor form
$ V_{k} \!=\! -i \, t_{fd}\; [ \sin(k_x)\!+\!\sin(k_y)\!+\!\sin(k_z) ] $.  The
three choices have zero, constant, and maximal $|V_{k}|$ on the
half-filled Fermi surface, respectively; and yield metallic,
insulating, and semi-metallic $f$-projected densities of states,
respectively, at $U=0$ and within a Fermi liquid phase.
We favor intersite $V_k$ due to the similarity of its 
metallic f-DOS to paramagnetic
local density results for the rare earths and to the character 
of their f-electrons the high pressure phases. However,
it is to be emphasized
that the essential results of this paper are independent of these
hybridization choices.

Our calculational approaches are 
dynamical mean-field theory (DMFT)\cite{DMFT,DMFT2,DMFTPAM}
and determinant quantum Monte Carlo (QMC)\cite{DET,QMCPAM}, carried out
for a three-dimensional simple-cubic lattice in the thermodynamic limit
and for a $4^3$ periodic lattice, respectively.
DMFT can capture thermodynamic phase transitions while
the quantitative agreement with QMC
provides compelling evidence its accuracy.

Within DMFT, both the f-electron part of the PAM and the HM map onto the same 
single site problem in which the Green function is calculated
from a dynamical mean field $\Sigma_f(\omega)$\cite{DMFT,DMFT2,HIRSCH}.
The difference between  PAM and HM
is the way  $\epsilon^f_{k}$ and $V_{k}$ enter 
the self-consistency  condition, i.e., 
the ${k}$ integrated Dyson equation
\begin{equation}
G_f(\omega) = 
\frac{1}{(2 \pi)^3} \int d^3 k \; \frac{1}
{\omega  - \Sigma(\omega) -  \epsilon^f_{k} - 
\frac{{V_k}^2}{\omega  - \epsilon^d_{k} }},
\label{dyson}
\end{equation}
where  $V_k=0$ and $\epsilon^f_{k}=0$ for the HM and PAM, respectively. 
In the DMFT equations the d-electrons of the PAM,
 which enter only quadratically, have been integrated out yielding an 
effective f-electron  problem with  retarded potential
${{V_k}^2}/(\omega  - \epsilon^d_{k} )$. 
One expects the different potentials 
to be reflected in quite distinct physical properties of the two models.

The spectral function $A_f(\omega)$ 
of the  f-electrons is shown in Fig.1. At 
small hybridizations $t_{fd}$, the PAM spectrum consists of two peaks
at $\pm U/2$ which corresponds to a phase with localized f-electrons.
At larger $t_{fd}$, an additional Abrikosov-Suhl resonance develops at 
the Fermi energy which is
not gapped for the PAM with intersite hybridization.
 Thus, the f-electrons are itinerant.
The striking similarity to
the spectral function of the HM (inset)
calls for a 
more quantitative comparison of  the crossover between itinerant and localized f-electrons which is 
driven by an increase of $U_f/t_{fd(ff)}$. Such a detailed comparison
is the subject of this paper.

\begin{figure}[p] 
\unitlength1cm 

\begin{center}
\epsfig{figure=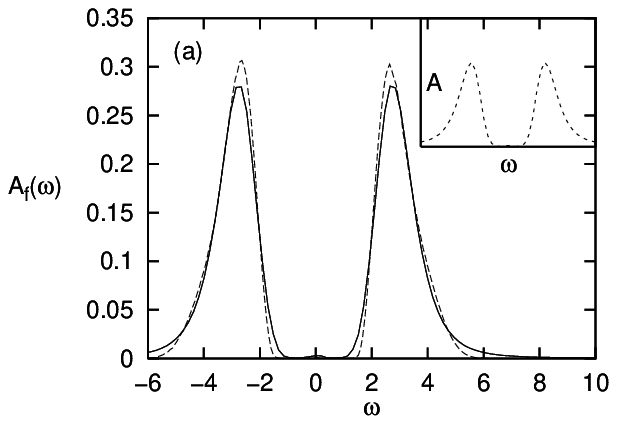,width=6.7cm,angle=0}

\epsfig{figure=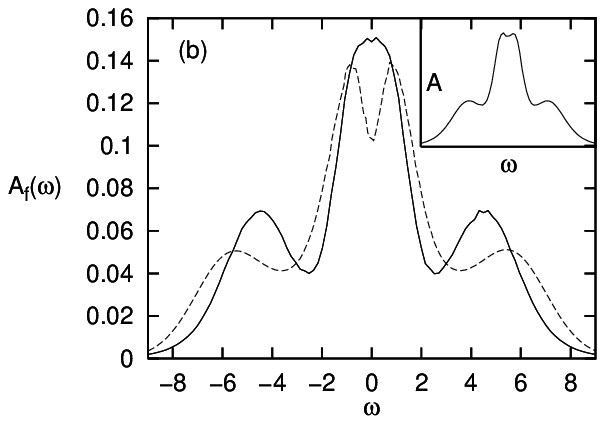,width=6.7cm,angle=0}

\end{center}

\caption{
The spectral function of the PAM at $U_f=6$ and  $T=0.2$
with an intersite hybridization
 of  (a)
$t_{fd}=0.3$ and (b) $t_{fd}=1$, respectively, 
as obtained by
DMFT (solid line) and QMC (dashed line).
With the 
exception of a small AF splitting of the central resonance 
found in finite-d simulations at  $t_{fd}=1$ the agreement between 
DMFT and QMC is excellent.
Insets: HM (DMFT) 
at  $U_f=6$, $T=0.2$, 
$t_{ff}=0.3$ and $t_{ff}=1$, respectively.
}
\end{figure}
In the self energy $\Sigma_f$ (Fig.2a), the crossover \cite{Note1} shows up as 
an abrupt change of the
behavior with $t_{fd}$:
for large $t_{fd}$ Fermi liquid behavior with
$\mbox{Im} \Sigma_f (i \omega_n) \propto  \omega_n$
at low (Matsubara) frequencies occurs, while at small $t_{fd}$
an  insulating-like behavior with 
diverging $\mbox{Im} \Sigma_f$ is observed.
From the $\Sigma_f$ data one can calculate a 
quantitative measure for the
crossover from itinerant to localized f-electrons: the 
quasiparticle weight of the  Abrikosov-Suhl resonance
$Z_f\!=\!(1\!-\!\partial \Sigma_f (\omega)/\partial \omega|_{\omega=0})^{-1}$
(calculated via  finite differential quotient $(1\!-\!\mbox{Im}
\Sigma_f(i\omega_0)/\omega_0)^{-1}$). Just  as at the Mott
transition of the HM \cite{BRINKMANRICE,GEBHARD,DMFT,SCHLIPF,NOAK},
$Z_f$ vanishes  with decreasing $t_{fd}/U_f$ for the PAM. 
This behavior of the PAM
is astonishingly similar to that seen in the HM
and is insensitive to the choice of hybridization
as is seen in Fig.~2b.

While the important similarity between the two models lies in the
presence or absence of the central resonance as a whole (measured by
$Z_f$), the PAM hybridization can determine finer details such as whether this
resonance may be split by a gap.  Consider the charge compressibility
$\kappa_f\!=\!\partial n_f/\partial \mu$ shown in Fig.~2c.  Nonzero
$\kappa_f$ indicates a metallic phase, while $\kappa_f$ is
exponentially small in $1/T$ within a gapped insulating phase.  At
small $t_{fd(ff)}$, all $f$ spectral functions show insulating behavior
as there are no central resonances lying between the two Hubbard
bands.  These resonances form in all cases at larger $t_{fd(ff)}$,
however, they are split by insulating and semimetallic hybridization gaps
for the onsite and odd-parity cases, respectively, as is evident in
Fig.~2c.
\begin{figure}[p] 
\begin{center}
\unitlength1cm

\hspace{-.3cm}{\psfig{figure=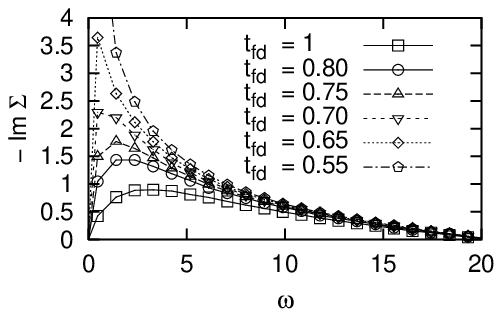,width=7.cm,angle=0}}

\hspace{-.351cm} {\psfig{figure=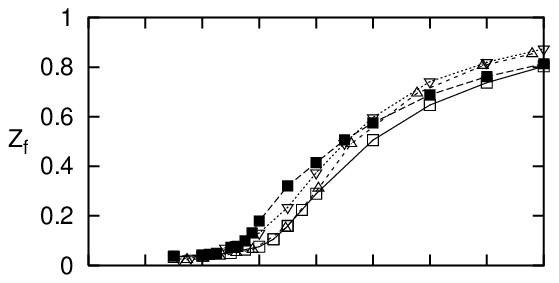,width=7.5cm,angle=0}}
\vspace{-1.3610cm}

\hspace{-.58cm}{\psfig{figure=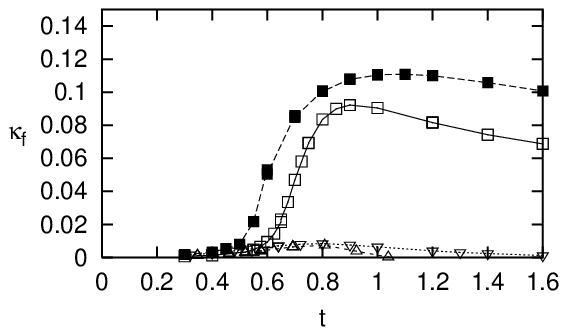,width=7.74cm,angle=0}}

\vspace{-11.52cm}

\hspace{-8.595cm}(a)

\vspace{4.cm}

\hspace{-8.595cm}(b)

\vspace{2.6cm}

\hspace{-8.595cm}(c)

\vspace{3.7cm}
\end{center}
\caption{
(a) (Matsubara) frequency dependence of the imaginary part of 
the self energy of the PAM with intersite hybridization.  
The behavior for the HM  (not shown) 
is very similar.
(b) Quasiparticle weight $Z_f$
as a function of $t=t_{fd}$
and $t=t_{ff}$ for the PAM and HM, respectively
($\Box$ : PAM with intersite,
$\vartriangle$: PAM with onsite, and
$\triangledown$: PAM with odd-parity intersite hybridization;
$\blacksquare$: HM). 
(c) Electronic compressibility $\kappa_f=\partial n_f/\partial \mu$ 
as a function of $t$.
All results are calculated by DMFT at $U_f=6$ and $T=0.15$ ($\vartriangle$, $\triangledown$: $T=0.07$). 
\label{qpweight}\\
} 
\end{figure}

While the (unscreened) local moment $\langle m_z^2 \rangle = \langle
  (n_{\uparrow} - n_{\downarrow})^{2} \rangle$  shown in  Fig.~3
  decreases with increasing hybridization, it is still quite
  substantial through the HM and PAM crossovers of Fig.2, marked by
  arrows in Fig.3. This smooth behavior of the local moment from weak
  to strong coupling in {\it both} models is insufficiently
  appreciated in the discussions of the distinctions between the 'Mott'
  and 'Kondo' scenarios for the volume collapse transition, where mean
  field treatments of the correlations lead to artificially abrupt
  changes in the local moment.  Despite the finite moments, screening
  effects lead to Pauli like low-T susceptibilities at large
  $t_{fd(ff)}$.
\begin{figure}[p] 

\begin{center}
\unitlength1cm
{\psfig{figure=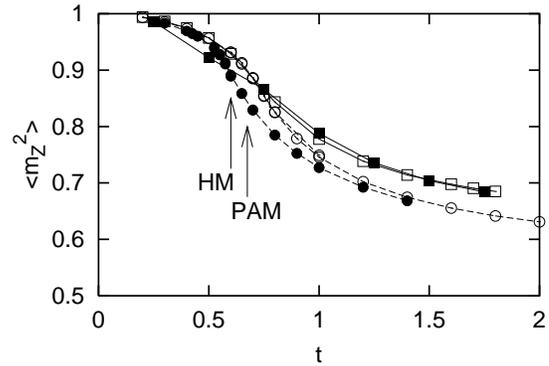,width=7.2cm,angle=0}}

\end{center}

\caption{ 
The square of the local moment is shown as a function of
$t=t_{fd}$ and $t=t_{ff}$  for the PAM with intersite hybridization 
(open symbols) and 
HM (closed symbols), respectively, at $U_f=6$ and $T=0.15$
(circles: DMFT, squares: QMC) [$T=0.5$ for the QMC HM to avoid antiferromagnetic ordering]. 
In neither case does the moment change
remarkably in the vicinity of the transitions of Fig.~2
(arrows).}
\end{figure}

The $T$-$U_f$ phase diagrams which result from consideration of the 
self-energy, quasiparticle weight,
and magnetic susceptibility are shown in Fig.~4.
Both the HM and PAM
have a low temperature phase of antiferromagnetic order
with a maximum in the N\'eel temperature  $T_N$   
at intermediate $U_f/W$.
 At a slightly
larger $U_f/W$, a second transition,
i.e., the  disappearance of the quasiparticle peak,
is observed within the  paramagnetic phase 
(if antiferromagnetism is frustrated). This
is the Mott transition of the HM and, 
from the Kondo point of view, the
crossing of the Kondo temperature $T_K(t_{fd})$.
Apart from the temperature and hybridization scales,
the phase diagrams of HM and PAM are remarkably similar
at finite $T$.  It is an open question at present whether such 
similarities will persist down to zero temperature.

  The present paper has shown a remarkable similarity between the HM
   and the f-electron system of the PAM which is indeed -- after
   integrating out the d-electrons -- like a HM with a more
   complicated retarded potential.  Novel numerical results on the
   spectral functions, charge compressibilities and local moments show
   that despite its more complicated potential, the PAM undergoes (at finite
   temperatures and half-filling)  a
   crossover from itinerant to localized f-electrons similar to the Mott
   transition of the HM.  While this similarity is closest for the
   PAM with intersite hybridization
   which describes metallic f-electrons at large $t_{fd}$ the
   vanishing of the central resonance does not depend on the 
   hybridization choice.
   This crossover and the transition
   towards  antiferromagnetic order yield phase diagrams which
   have the same topology for both models. Accordingly, the present paper
 suggests that the physics underlying
  the  Mott \cite{JOHANSSON} 
\begin{figure}[p] 
\begin{center}
\unitlength1cm 
{\psfig{figure=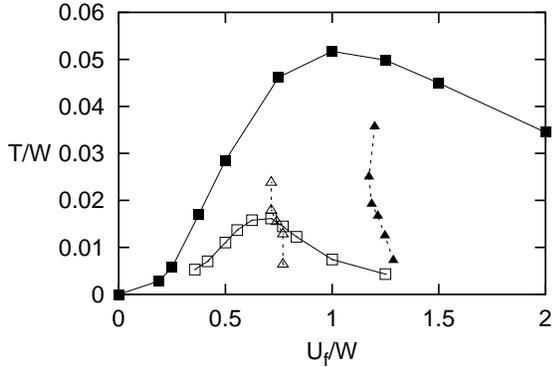,width=7.2cm,angle=0}}
\end{center}

\caption{
$T$-$U_f$ phase diagram of the Hubbard model with a semi elliptic DOS
(filled symbols; reproduced from \protect
\cite{ULMKE,SCHLIPF})
and PAM with intersite hybridization (open symbols) within DMFT.
Here $W$ is the band width
of the Bethe DOS and $12 \; t_{fd}$ for the PAM, respectively.
Squares:  N\'eel temperature; triangles: 
Mott transition/Kondo temperature (the dashed lines end
at higher $T$ as the crossover becomes too gradual in
$U_f/W$ to allow a precise location).
}

\end{figure}
\noindent
 (HM) and the Kondo volume collapse
  scenarios \cite{ALLEN} (PAM or its impurity approximation) for the
  volume collapse transitions in f-electron
  metals\cite{BENEDICT,JCAMD} may be similar and {\em not}
  incompatible.
While the Mott transition is continuous at the temperatures considered, 
the rapid change in the correlation energy, when combined with the smooth 
contributions from other valence bands, might nevertheless still result 
in a Maxwell construction giving the thermodynamic instability manifested 
as the first-order volume collapse in Ce \cite{delp}. 

In regard to local density calculations, orbitally-polarized and self-interaction-corrected 
applications \cite{LDA} to the volume-collapse transitions resemble a
mean field solution of the HM, where the  transition
coincides with a coalescence of the two Hubbard bands and a simultaneous
loss of local moment.
The implication of the present paper is that
regardless of the balance between f-f (HM) and f-valence hybridization (PAM),
a {\it correlated} solution should show some Kondo-like
attributes in the collapsed phase, including an analog of the
three-peak local DOS,  persistence of the local moment,
and its screening as deduced from
the susceptibility.
Local density theory does
        not show such Kondo-like features in the collapsed phase,
        however, it does nevertheless yield excellent
        structural-dependence of the total energy \cite{LDA2}.

{\small
We acknowledge valuable discussions with N. Bl\"umer, P. van Dongen,
M. Jarrell, and D. Vollhardt.
Work  was supported in part by a DOE ASCI grant, the DAAD, 
the LLNL MRI, 
the U.S. DOE Contract No.~W-7405-Eng-48, and
NSF grant DMR9704021.}


\begin{references}



\bibitem{GEBHARD}
F.~Gebhard, {\it The Mott metal-insulator transition},
Springer-Verlag, Berlin 1997.

\bibitem{DMFT}
A. Georges {\it et al.}, 
Rev. Mod. Phys. {\bf 68} 13 (1996).

\bibitem{HUBBARDIII}
J.~Hubbard, Proc.~Roy.~Soc.~London {\bf A 281}, 281 (1964).

\bibitem{NEWNS}
D.~M.~Newns and N.~Read,
Adv.~Phys. {\bf 36}, 799 (1987).

\bibitem{HUSCROFT} 
C.~Huscroft {\it et al.},   Phys.~Rev.~Lett.~{\bf 82}, 2342 (1999).



\bibitem{DMFT2}    
D. Vollhardt, in {\it Correlated Electron Systems},
V. J. Emery ed. (World Scientific, Singapore) 57 (1993);
Th. Pruschke {\it et al.}, 
Adv. Phys. {\bf 44}, 187 (1995);

\bibitem{QMCHM}
M. Ulmke {\it et  al.},
Phys.~Rev.~{\bf B 54},  16523 (1996).


\bibitem{JOHANSSON}
B. Johansson, Philos. Mag. {\bf 30}, 469 (1974);
B.~Johansson {\it et al.},
Phys.~Rev.~Lett. {\bf 74}, 2335 (1995).

\bibitem{ALLEN}
J.W. Allen and R.M. Martin, Phys. Rev. Lett. {\bf 49} 1106, (1982);
J.W. Allen and L.Z. Liu, Phys. Rev.  {\bf B 46} 5047, (1992);
M. Lavagna {\it et al.},
Phys.  Lett. {\bf 90A}, 210 (1982); J. Phys. F {\bf 13}, 1007 (1983).


\bibitem{BENEDICT}
See, e.g., U. Benedict {\it et  al.}, Physica {\bf 144B}, 14 (1986).

\bibitem{JCAMD}
A.~K.~McMahan {\it et al.},  J. Comput.-Aided Mater. Design {\bf 5},
131 (1998).



\bibitem{DMFTPAM}
DMFT studies of the onsite hybridization include: 
M.~Jarrell {\it et al.},
 Phys.~Rev.~Lett.~{\bf 70}, 1670 (1993); 
M.~Jarrell, Phys.~Rev.~{\bf B 51},
7429 (1995); A.N.~Tahvildar-Zadeh {\it et al.},
Phys.~Rev.~{\bf B 55}, 3332 (1997);  
M. Rozenberg, Phys. Rev. {\bf B 52}, 7369 (1995).

\bibitem{DET}
R. Blankenbecler {\it et al.}, 
Phys. Rev. {\bf D 24}, 2278 (1981).

\bibitem{QMCPAM}
QMC studies of the onsite hybridization include: 
J. Bon\u{c}a and J.E.~Gubernatis,
Phys.~Rev.~{\bf B 58}, 6992 (1998); Y.~Zhang and J.~Callaway,
Phys.~Rev.~{\bf B 38}, 641 (1988);
M.~Vekic {\it et al.}, 
Phys.~Rev.~Lett.~{\bf 74}, 2367 (1995).    


\bibitem{HIRSCH}
The DMFT single site problem is solved numerically following 
J.~E. Hirsch and R.~M. Fye,
\newblock Phys. Rev. Lett. {\bf 56}, 2521 (1986).

\bibitem{Note1}
While the transition observed is not first-order (at the temperatures 
considered) it is very difficult to decide  numerically
whether it is a second-order phase transition or non-critical.
We use the word ``{crossover}''  to expresses our expectation 
that it is non-critical without excluding the possibility that it
is second-order. 



\bibitem{BRINKMANRICE}
W.F.~Brinkman and T.M.~Rice, Phys. Rev. {\bf B 2,} 4302 (1970).



\bibitem{SCHLIPF}
J.~Schlipf {\it et al.}, 
 Phys.~Rev.~Lett. {\bf 82}, 4890 (1999).

\bibitem{NOAK}
 
R.M.\ Noack and F.\ Gebhard, Phys.~Rev.~Lett. {\bf 82}, 1915 (1999);\
R.~Bulla,  Phys.~Rev.~Lett. {\bf 83}, 136 (1999).

\bibitem{ULMKE}
N.~Bulut {\it et al.}, Phys.~Rev.~Lett. {\bf 72}, 705 (1994);
M.~Ulmke  {\it et al.},
Phys.~Rev. {\bf B 51}, 10411 (1995).

\bibitem{delp} We obtain energy difference curves for the present
    DMFT PAM results quite similar to the QMC results in Fig.2(a) of
    \cite{HUSCROFT}, with slope-changes near $t_{fd} \sim 0.6$ of
    $\partial \Delta F/\partial \ln{t_{fd}} \sim 0.4$ eV for {\it both}
    on-site and even-parity nearest-neighbor hybridizations (the only
    two tested), which as noted in that paper are in order of magnitude
    agreement with experiment.

\bibitem{LDA}
O. Eriksson {\it et  al.}, Phys. Rev.  {\bf B 41} 7311 (1990); A. Svane
{\it et  al.}, Phys. Rev.  {\bf B 56} 7143 (1997); Z. Szotek {\it
et al.}, Phys. Rev. Lett. {\bf 72} 1244 (1994); A.~Svane, Phys. Rev.
Lett. {\bf 72} 1248 (1994); 
Phys. Rev.  {\bf B 53} 4275 (1996).
\bibitem{LDA2} P. S\"oderlind, Adv. Phys. {\bf 47}, 959 (1998).








\end{references}
\end{document}